\documentclass[fleqn,usenatbib]{mnras}

\usepackage[T1]{fontenc}

\DeclareRobustCommand{\VAN}[3]{#2}
\let\VANthebibliography\thebibliography
\def\thebibliography{\DeclareRobustCommand{\VAN}[3]{##3}\VANthebibliography}

\usepackage{graphicx}	
\usepackage{amsmath}	
\usepackage{amssymb}	

\title[Stellar phylogenies]{Using heritability of stellar chemistry to reveal the history of the Milky Way}

\author[Jackson \& Jofr\'e et al.]{
Holly Jackson,$^{1}$
Paula Jofr\'e,$^{2}$\thanks{E-mail: paula.jofre@mail.udp.cl}
Keaghan Yaxley,$^{3}$ 
Payel Das,$^{4}$ 
Danielle de Brito Silva,$^{2}$ 
and Robert Foley$^{3}$
\\
$^{1}$Department of Electrical Engineering and Computer Science, Massachusetts Institute of Technology, 50 Vassar Street, Cambridge,
MA 02139, USA\\
$^{2}$N\'ucleo de Astronom\'ia, Universidad Diego Portales, Ej\'ercito 441, Santiago, Chile\\
$^{3}$Leverhulme Centre for Human Evolutionary Studies, Department for Anthropology and Archaeology, University of Cambridge, CB2 1QH, UK \\
$^{4}$Department of Physics, University of Surrey, Stag Hill, University Campus, Guildford, GU2 7XH, UK
}

\date{Accepted XXX. Received YYY; in original form ZZZ}

\pubyear{2020}

\begin{document}
\label{firstpage}
\pagerange{\pageref{firstpage}--\pageref{lastpage}}
\maketitle

\begin{abstract}
Since chemical abundances are inherited between generations of stars, we use them to trace the evolutionary history of our Galaxy. We present a robust methodology for creating a phylogenetic tree, a biological tool used for centuries to study heritability. Combining our phylogeny with information on stellar ages and dynamical properties, we reconstruct the shared history of 78 stars in the Solar Neighbourhood. The branching pattern in our tree supports a scenario in which the thick disk is an ancestral population of the thin disk. The transition from thick to thin disk shows an anomaly, which we attribute to a star formation burst. Our tree shows a further signature of the variability in stars similar to the Sun, perhaps linked to a minor star formation enhancement creating our Solar System. In this paper, we demonstrate the immense potential of a phylogenetic perspective and interdisciplinary collaboration, where with borrowed techniques from biology we can study key processes that have contributed to the evolution of the Milky Way.
\end{abstract}

\begin{keywords}
Milky Way evolution -- phylogeny -- stellar abundances
\end{keywords}



\section{Introduction}
After the publication of Darwin’s \textit{the Origin of Species} in 1859, 
it took almost a century for DNA to be recognised as the mechanism for biological inheritance. It is a molecule that allows the traits of an organism to be passed from one generation to the next. Yet without any knowledge of DNA, Darwin understood that heritability underpinned descent with modification (see App.~\ref{descent}), which in turn underpinned evolution. He depicted the patterns of descent among organisms as an evolutionary tree. If there is some heritability in a system, then a tree of descent is an extremely apt model. Today, in all branches of biology,  trees -- now more generally known as phylogenies -- are a major tool for analysing evolutionary histories.

At first sight, it might seem that the underlying principle of heritability necessary for a phylogenetic analysis does not occur in galaxy evolution. After all, they have no DNA or genes. However, heritability does play a role in the chemical evolution of galaxies. The stars forming and dying in galaxies are both carriers of chemical information and responsible for the modification and evolution of galactic chemical composition. Indeed, stars are the main producers of chemical elements heavier than helium in the Universe \citep{Burbidge1957, kobayashi2020origin}. In each star formation episode, stars of a wide range of masses are born. The massive stars die quickly and eject new heavy elements into the existing interstellar medium (ISM), hence modifying the galactic chemical composition. This new material collapses to form gas clouds, which form new generations of stars. These stars inherit the chemical composition of the dead stars \citep{Chiappini1997,matteucci2012chemical, andrews2017}. 

Low-mass stars live longer, and their atmospheres preserve the chemical composition of their ancestral gas clouds, hence serving as fossil records for chemical evolution studies \citep{Freeman2002}. These long-lived, low-mass stars form the bulk of stellar observations in the Milky Way. With Gaia, we have now more than one billion positions and motions of individual stars \citep{Gaia2018}, which can be combined with spectral data from several spectroscopic surveys \citep[see e.g. recent review of][]{Jofre19}. This allows the study of the distribution of chemical abundances across the Galaxy \citep{wheeler2020}, which combined with information on stellar ages where available, can provide crucial information on how chemical patterns have evolved with time \citep{ness2016, Buder2019}.   

Despite the wealth of current data, combining the multidimensional information of individual ages, motions, positions and different chemical abundances of millions of stars into a few logical final visual products that paint the history of our Milky Way is very challenging. This has been a barrier to addressing some of the long-standing questions about the creation of the chemical elements and how they encode the formation and evolution of the Milky Way.

Attempts to reduce the dimensionality of the datasets have been published recently \citep[e.g., ][]{anders2018, Garcia-Dias2019, price-jones2020}, showing that stars from different groups indeed trace different chemical evolution histories. Understanding of how the stars within each group and between groups are connected within the evolutionary context of the Milky Way as a whole is still missing.

Combining multidimensional datasets to reconstruct history is a challenge in most evolutionary studies. Long-term history must be reconstructed from sparse signals from the past -- fossils in the case of biology, the chemical patterns of stars in the case of astronomy. These fossil records contain information that span a wide range of dimensions. If that information is heritable, then it is possible to use a phylogenetic approach. Hence, we can borrow the phylogenetic approach from biology to reconstruct galactic evolution.

In any evolutionary analysis, there are two components -- one, the actual phylogeny of the constituent lineages, and two, the context, which is assumed to influence the shape of the tree. The former is reconstructed from the heritable traits of the organisms, and the latter by the environment shaping the phylogeny (for example, climate change in the case of species on Earth). In galaxy evolution, the traits or variables contributing to heritable component are those encoding the chemical pattern of the stars. Variables that reflect the context include those that describe the dynamical `environment' of the star, such as interactions with the bar and spiral arms and/or the galaxy's external environment, which could lead to changes in positions through processes such as radial migration and heating.

In this paper, we assemble data on the chemical composition of low-mass stars  and organise these into a format amenable to phylogenetic analysis, so that stars can be linked in a tree that reflects shared ancestry. Using this approach, we build a tree of nearby stars covering a wide range of ages, with the aim of revealing evolutionary pathways in the Milky Way. We then use the context, inferred from the ages and motions of the stars and their relationship to each other, to identify the processes of Milky Way formation, from its ancient to its present form. 

Other standard astronomical approaches may achieve similar conclusions to what is presented here, but usually with a larger sample of stars and with the consideration of models. Here we will show that the advantage of a phylogenetic approach lies in the significant information about shared histories that can be empirically uncovered with a small sample of stars. Firstly, the overall structure of the tree can reflect particular processes (in biological evolution, for instance, the difference between a tree with a burst of speciation (see App.~\ref{speciation}) suggests a very different process from one where branching occurs evenly spread throughout the tree). Secondly, linking branching events to ages can inform the timing of events; and lastly, trees are seldom perfect, and the anomalies and outliers can often indicate important parts of the history worthy of further investigation. 

In Section \ref{sect:data} we describe the stellar data used for this work and in Section~\ref{sect:methods} we explain a new robust methodology to create a phylogenetic tree with stellar chemical data. Our results are presented in Section ~\ref{sect:results} which are discussed and interpreted in terms of Galactic history in Section ~\ref{sect:discussion}. We finish the paper with our Conclusions in Section ~\ref{sect:conclusions}

\section{Data}\label{sect:data}
Below we introduce the sample of stars for which we create a phylogenetic tree, and the abundance ratios we use to depict their chemical patterns. The stellar data used for our analysis (traits, ages and kinematics) can be found in the online table.

\subsection{Sample}\label{ssect:sample}
We chose stars that have very similar spectra 
to the Sun (i.e. they are `solar twins'), which allows chemical abundances and ages to be derived relative to the Sun, achieving the highest accuracy and precision possible in stars \citep[see reviews of][for further discussions]{Nissen&Gustafsson2019, Jofre19}. This incurs a bias in our perspective of the Milky Way, as the stars are similar in iron content and are all close to us. However, they can still have a significant spread in ages (from 0 to 10 Gyr) and abundance ratios of different chemical elements can differ up to 1~dex. Thus they still serve as snapshots of different evolutionary epochs \citep[see e.g. ][]{Nissen2017, Nissen2020}. 
 
We use a sample of solar twins whose chemical abundances and ages have been determined and published  
by \cite{Bedell2018}.  
The sample comprises 79 stars (including the Sun) and abundances are measured for 30 elements, in addition to precise ages. 
We exclude star HIP64150 for having peculiar chemical abundances, probably due to mass-transfer episodes with a companion \citep[see details in][]{Bedell2018}, leaving a final sample of 78 stars. 

The stellar abundances were derived by the respective authors following standard procedures of stellar spectroscopy from the high-resolution HARPS instrument. They determined line-by-line differential abundances with respect to the Sun. To do so, they measured equivalent widths from lines that are clean and not saturated in the star's spectrum and that of the Sun. Using 1D-LTE atmosphere models they determined the abundance from the equivalent widths using the linear part of the curve of growth. The median uncertainty in the abundance measurements is 0.008 dex corresponding to the line-to-line scatter. Ages were determined using standard isochrones, also differentially with respect to the Sun. Details about this procedure are found in \cite{Bedell2018} and extensive discussions and applications of the differential abundance determination method can be found in \cite{Nissen&Gustafsson2019}. 

Kinematic information is taken from the HARPS data and the second data release (DR2) of Gaia \citep{Gaia2018}. The HARPS spectra provide radial velocities of the stars and Gaia DR2 provides sky positions, parallaxes, and proper motions. As the stars are all very nearby, their distances are well approximated by the inverse of the parallax.

The kinematics are analysed in terms of actions and birth radii. Actions are integrals of motion, always the same along each orbit \citep{binney2011galactic}. $J_r$ is the radial action and measures the range of motion in the Galactic plane. $J_z$ is the vertical action and measures the range of motion above and below the galactic plane (the $z$ coordinate). $L_z$ is the $z$ component of angular momentum and gives an approximate radius in the galactic plane for the orbit. The actions can be combined ($\frac{J_r+J_z}{|L_z|}$) to give a measure of the eccentricity of the orbit. Using dynamical models, the motions of the stars can be followed backwards in some assumed gravitational potential for the time they have existed to predict the radii at which they are born ($R_b$). Due to theoretical uncertainties in the Milky Way potential and stellar age, as well as uncertainties in the measurements of stellar motions, the birth radius becomes quite uncertain for stars older than 1 Gyr.  With this information, we determine the three actions (the radial action $J_r$, the vertical action $J_z$, and the $z$ component of angular momentum) using the action finding-package in AGAMA \citep{Vasiliev19} and assuming the most probable Milky Way gravitational potential found in \cite{Mcmillan2017}.

\subsection{Chemical abundance ratios used as traits}\label{ssect:sample}
Since many chemical elements are produced at similar astrophysical sites, we have chosen traits that are evolutionary informative \citep{Jofre2020}. Adding traits that do not evolve with time or that do not vary significantly among stars only adds noise to any phylogenetic signal that may be in the data and therefore can be ignored.  
We select the abundance ratios presented in \cite{Jofre2020} to produce our tree. These correspond to  [C/Y], [C/Zr], [C/Ba], [C/La], [C/Ce], [C/Nd], [O/Y], [O/Ba], [Na/Sr], [Na/Y], [Na/Zr], [Na/Ba], [Na/La], [Na/Ce], [Na/Nd], [Mg/Sr], [Mg/Y], [Mg/Zr], [Mg/Ba], [Mg/La], [Mg/Ce], [Mg/Nd], [Al/Y], [Al/Ba], [Si/Zr], [Si/Ba], [Si/La], [Si/Ce], [Ca/Ba], [Sc/Sr], [Sc/Y], [Sc/Zr], [Sc/Ba], [Sc/La], [Sc/Ce], [Sc/Ba], [Ti/Y], [Ti/Ba], [Ti/Ba], [Mn/Y], [Mn/Ba], [Co/Ba], [Ni/Sr], [Ni/Y], [Ni/Zr], [Ni/Ba], [Ni/La], [Ni/Ce], [Cu/Sr], [Cu/Y], [Cu/Zr], [Cu/Ba], [Cu/La], [Cu/Ce], [Zn/Ba]. 

Some of our heritable traits have been discussed previously in the context of ``chemical clocks'' \citep{dasilva2012, Nissen2015, 2019A&A...624A..78D}. As recently extensively discussed in e.g. \cite{Casali2020}, not all ``chemical clocks" hold for all stellar samples, and it is yet to be seen how their dependency with age changes accross the Galaxy and parameter space. The seemingly  non-universality of these abundance ratios as a function of stellar age motivates us to treat them simply as heritable traits instead of a measure of time like a clock \citep[see further discussions in][]{Jofre2020}.   

Only the chemical abundances are considered as herirable traits to build our trees, since the only heritable information for chemical evolution are the chemical abundances.  Ages and kinematics are only used to help us with the interpretation of the tree in terms of Galactic history but should not be used to construct phylogenetic trees.

\section{Methods}\label{sect:methods}

In this Section, due to the interdisciplinary nature of our work, we give a detailed account of constructing phylogenetic trees, starting with a brief overview of the basics of phylogeny relevant to this study. In-depth explanations of the foundation of phylogenetics can be found in the literature \citep{Baum2005, FelsensteinBook, Felsenstein1988, Hall2004, Lemey2009}. We then explain the steps involved in generating a tree of stars using chemical abundances as traits. 

\subsection{Building phylogenetic trees of stars}
Phylogeny tells us which of the taxa (see App.~\ref{taxa}) we wish to compare (e.g. organisms, species, viruses, etc.) are most closely related. Phylogenetic trees visualize these relationships \citep{Baum2005, FelsensteinBook, Felsenstein1988, Hall2004, Lemey2009}. Phylogenetic methods have already been applied beyond evolutionary biology, to model the evolution of language and other human cultural activities \citep{Gray2009, Retzlaff2018}, e/g/, they are not restricted to modelling the evolution of genes only. In the application of phylogenetic methods to reconstruct the evolutionary history of the Galaxy, the stars are the taxa, as they carry their evolutionary information in their chemical makeup. Even though these stars shine in the sky today, their chemical makeup represents that of the gas cloud at the time and place they were born. Therefore, our phylogenetic tree tracks the relationships of these stellar fossils to study the evolving insterstellar medium of the Galaxy.

Other astronomy studies have also attempted to apply phylogenetic methods to chemical abundance data in stellar samples. For example,  \citet{BlancoCuaresma2018} used the concept of phylogeny from biological clustering methods to classify stars based on their chemical similarities. \cite{Jofre2017} applied phylogenetic methods to stellar data with the aim of revealing the evolutionary history of the galaxy as a proof of concept. 

Our goal here is not to order the stars in similarities and differences, but to use the principle of descent with modification (see App.~\ref{descent}) and chemical heritability to reveal the evolution of the interstellar medium from which these stars formed. Hence, we must first order the stars in similarities and differences in their chemistry similar to \cite{BlancoCuaresma2018}. To do so, we use the methods detailed in our previous paper \citep{Jofre2017} as a baseline but have developed a far more robust pipeline, based on phylogenetic studies in evolutionary biology. It contains three steps: (1) encoding evolutionary traits, (2) tree generation, and (3) evaluating tree robustness.

\subsection{Encoding evolutionary traits}\label{sec:distmat}
Since existing modern phylogenetic tools are intended for biological applications, they are specially designed to take molecular (DNA) data as a discrete sequence of thousands of nucleotides, \citep{Drummond2007, Hall2013, Maddison2009}. In contrast, the evolutionary information of stars (i.e. stellar ``DNA'') is encoded in their elemental abundance ratios, and is continuous. Additionally, as opposed to thousands of genes, stellar data contains only tens of elemental abundance ratios per taxa.  
We still do not fully understand which chemical abundance ratios best trace chemical evolution for all stellar populations because of the variety of dependencies in the physical processes of galaxy assembly and the uncertainties in stellar yields \citep[see e.g][]{matteucci2012chemical, kobayashi2020origin}. Nor do biologists fully understand how genes evolve in every species in every place since the beginning of life on Earth. Phylogenetic trees are the tools designed to help understanding how all of this combines. 

Before evolutionary biologists began using nucleotides as traits, they too presented heredity traits as a continuous measurement. As a result, they developed methods, which use distance matrices to quantify similarities between the taxas' traits \citep{Lemey2009}. Recent advances in the field demonstrate that although DNA is the best feature for studying life on Earth, phylogenetic trees based on nucleotides alone can not represent the complete history of species. For example, DNA sequences are not available for all species (especially those that are extinct). This has profound implications in e.g. predicting the existence of very distinct species in the past, such as some kinds of dinosaurs \citep{bromham2016introduction}. Therefore, several methods relying on other traits measured from e.g. fossil evidence, play a fundamental role to constrain molecular evolution. That motivates us to use such methods based on distance matrices to encode our continuous traits (the stars’ elemental abundances). 

The distance matrix contains the pairwise chemical distance of all stars in our sample. We experimented with using Euclidean, Manhattan, and Minkowski distance methods of each of our abundance ratios. The choice of distance metric had a minimal impact on the differences between the final results and conclusions.  
Our final implementation used the Euclidean distance.

\subsection{Tree generation}\label{sec:treegen}

While the ideal method to find a phylogenetic tree for a set of data would be to evaluate every possible tree, this quickly becomes computationally intractable.  For $n \geq 2$ taxa, $\frac{(2n-3)!}{2^{n-2}(n-2)!}$ possible rooted trees (see App.~\ref{root}) exist \citep{Felsenstein1978}. This is on the order of 
$10^{139}$ possibilities for the 78 stars in our dataset.

The computational complexity motivates the use of one of the many simplified distance methods for tree generation. The most basic ones are the unweighted pair-group method with arithmetic mean (UPGMA), Fitch-Margoliash, and neighbor-joining (NJ) \citep{Sokal1958, Fitch1967, Saitou1987} algorithms. After more than a century of research on constructing phylogenetic trees, modern evolutionary biology studies tend to use more sophisticated phylogenetic methods that use a Bayesian analysis to create and interpret evolutionary trees \citep{ FelsensteinBook, Lemey2009, Drummond2012}. However, these methods require the definition of a scientific model for trait evolution to be incorporated into a tree. We intend to develop this in future work.  



Each distance method makes several assumptions about the evolution of the taxa.  Several of these assumptions assume they are molecular clocks with a constant evolutionary rate of change (e.g. a universal rate of evolution). They  contradict our theories for chemical evolution, since we know that chemical evolution happens on different timescales at different places \citep{matteucci2012chemical, Casali2020}. The NJ  algorithm is the best method for our purpose, as it does not assume the taxa are molecular clocks \citep{Gascuel2006}. UPGMA assumes molecular clocks, and the Fitch-Margoliash method requires a complex correction for error when the evolutionary rates of the taxa differ \citep{Lemey2009, FelsensteinBook}.


%
The NJ algorithm is not only efficient, but produces reliable phylogenies, and has been compared to more sophisticated, probability-based phylogenetic methods \citep{atteson1997performance, Kuhner1994, Lemey2009, Mihaescu2009}. Furthermore, the NJ algorithm was employed in \cite{Jofre2017}, allowing us to make better comparisons of our findings with the previous ones. 
 
\subsection{Evaluating tree robustness}
While the output of the NJ algorithm might provide a viable empirical evolutionary tree for the data, the final tree should account for uncertainties in the trait measurements.  

To do this, phylogeneticists often apply a bootstrapping algorithm to incorporate known and/or random uncertainty into the data.  After generating many trees from hundreds to thousands of resampled datasets, the trees are combined into a final consensus tree, which represents the most stable phylogenetic result.  In this study, we implemente a parametric bootstrapping algorithm (i.e. Monte Carlo sampling) and the maximum-clade credibility algorithm.


Parametric bootstrapping involves running a Monte-Carlo simulation based on the distributions of the original data points \citep{FelsensteinBook}. Since we know the distributions from which our data points were drawn (i.e. we have the uncertainties in the abundance measurements), we can use parametric bootstrapping to resample the original data.  In the end, we generate 1000 phylogenetic trees based on random resamplings of the uncertainties in the abundance measurements, reflected in 1000 different distance matrices.

To synthesize our sample trees into a final, possible result, we use the maximum-clade-credibility algorithm \citep{Drummond2007}. The maximum-clade credibility algorithm  evaluates every clade (see App.~\ref{clade}) in each sample tree in order to find the tree whose clades have the highest support. When generating the final tree, the algorithm finds which tree from the sample trees has the highest overall support.

A popular alternative to maximum-clade credibility is the majority-rule consensus algorithm, which was employed in \cite{Jofre2017}. Majority-rule consensus synthesizes the most common bifurcations (those that occur at least 50\% of the time) in the posterior trees into a final output tree \citep{Margush1981}. However, the majority-rule consensus algorithm has few drawbacks. Firstly, the final tree is very dependent on the cutoff threshold.  Usually the value employed is 50\%, which would allow a branch with 51\% confidence to trump a branch with 49\% confidence in the final output \citep{Lemey2009}. This seems too arbitrary for our stellar data. 
Secondly, it is difficult to represent accurate branch lengths in the final majority-rule-consensus tree because each selected bifurcation may come from a separate topology \citep{Lemey2009}. Finally, the algorithm builds a new tree that may have never been sampled during bootstrapping, so the final consensus tree may not be one of the sample trees \citep{Cranston2007} and therefore might not represent any true phylogeny.


As a final step, we truncate short branches in the final tree to guarantee our branching pattern is not a product of uncertainties. The performance of NJ will be accurate if all the error values in the distance matrix are less than 1/2 the shortest branch length in the tree \citep{atteson1997performance}. 
This motivates truncation of branches shorter than $3 \sigma$ (0.024 dex) in the final tree, where $\sigma$ (0.008 dex) is the median uncertainty in the abundance measurements.

However, truncation will cause multifurcations, or {\it polytomies} (see App.~\ref{polytomy}), in the tree.  While polytomies are often considered signs of a faulty tree in evolutionary biology, in the stellar case, they may be appropriate.  Because there are so many uncertainties in astronomical measurements, having an unresolved branching pattern is better than having an incorrect one.  Polytomies can  hint towards interesting events happening in the past, as we shall discuss further. 


The trees are generated using the R packages {\tt phangorn} and associated libraries \citep{Schliep2011,Schliep2017} 

\begin{figure*}
\centering
\includegraphics[width=\textwidth]{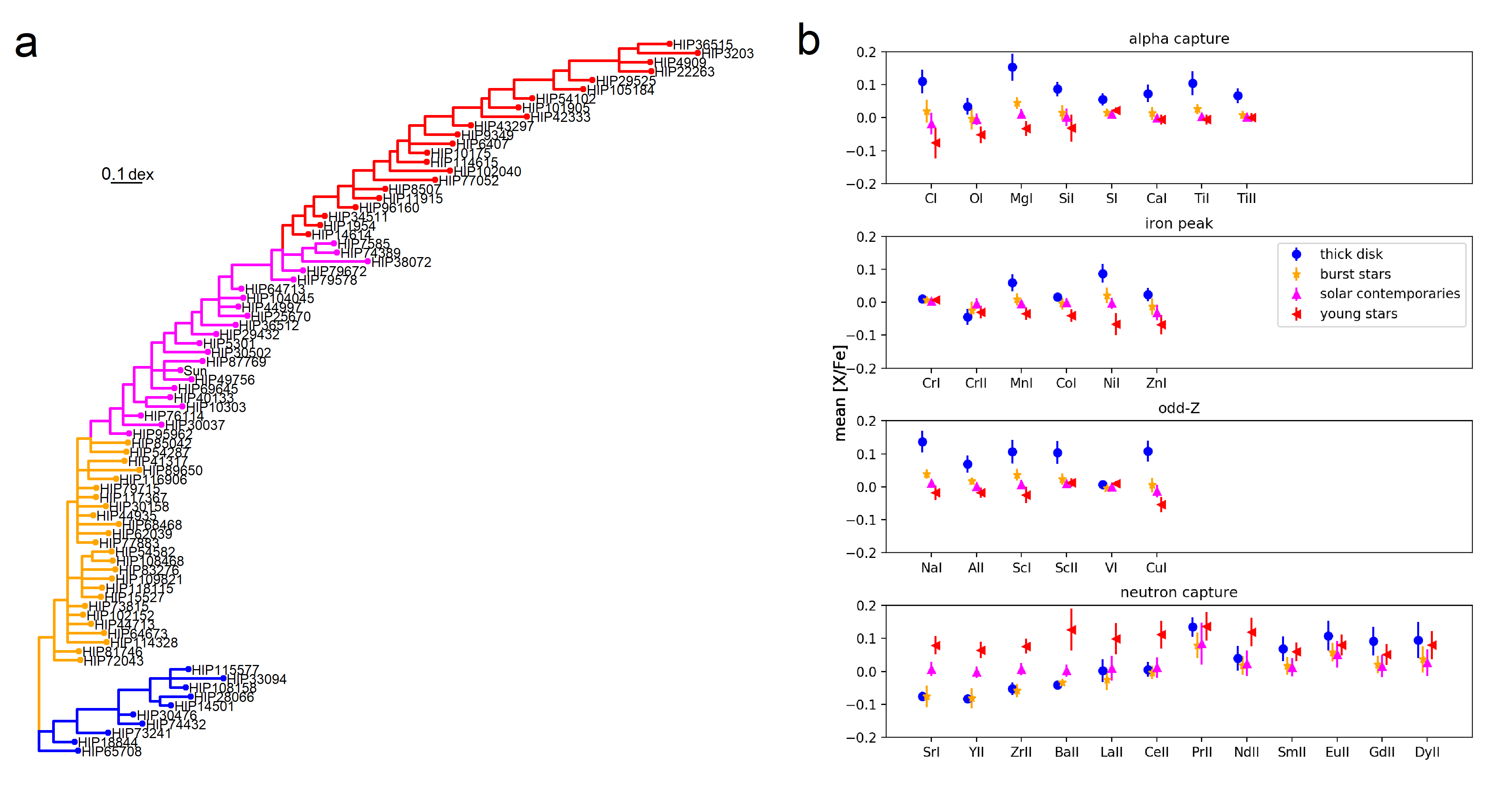}
\caption{(a) Classification of populations following their location in the tree. Blue: Thick disk, yellow: burst, magenta: Solar contemporaries and red: young. (b) mean abundances and standard deviation of stars in the group for all abundances
considered as a function of iron. Abundances are separated according to their nucleosynthesis process}
\label{fig:abundance_groups}
\end{figure*}

\section{Results}\label{sect:results}

Our resulting tree is shown in Fig.~\ref{fig:abundance_groups}a. The tree represents the hierarchical differences of pair-wise chemical Euclidean distances of each star included in our sample. To understand the different stellar populations in our tree better, we have coloured stars according to their position in the tree. This helps us to attribute the stars to astrophysically meaningful stellar populations.

\subsection{Stellar populations}\label{sect:stellar_pops}
\begin{figure}
\centering
\includegraphics[scale=0.4]{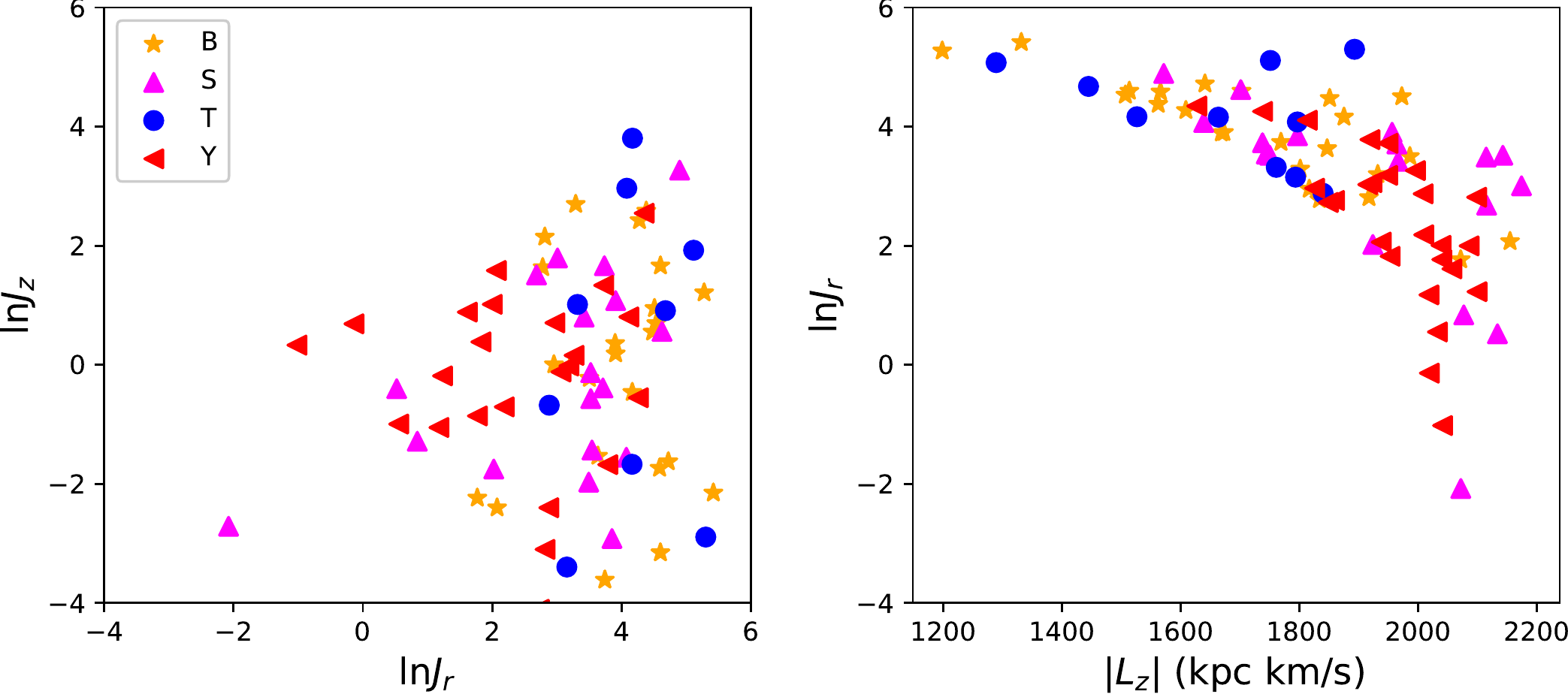}
\caption{Logarithm of the vertical action against the logarithm of the radial action shown on the left, and logarithm of the radial action against the absolute z-component of angular momentum on the right. The left plot shows that the T (blue circles) and B (orange stars) groups of stars generally have higher radial actions than other stars, and slightly higher vertical actions. The right plot shows that the T (blue circles) and B (orange stars) groups of stars generally have lower z-components of angular momentum. }
\label{fig:kinematics1}
\end{figure}

\begin{figure*}
\centering
\includegraphics[scale=0.6]{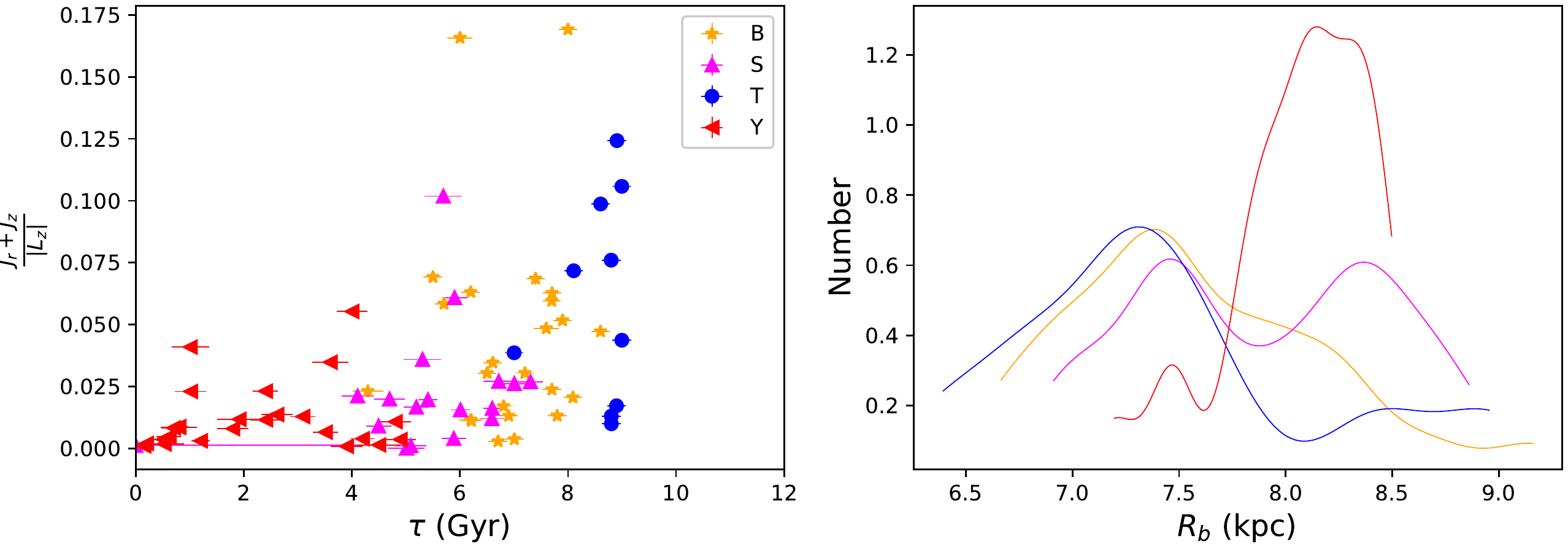}
\caption{A measure of eccentricity against age shown on the left, and the distribution of birth radii shown on the right. The left plot shows that older stars generally follow more eccentric orbits. The right plot shows that the T (blue circles) and B (orange stars) groups of stars generally have smaller birth radii.}
\label{fig:kinematics2}
\end{figure*}

To further examine the general properties of our stars we examine the groups in terms of their chemical abundances (Fig. \ref{fig:abundance_groups}b), age, and kinematic variables (Fig.~\ref{fig:kinematics1} and ~\ref{fig:kinematics2}), respectively. We examine them in terms of [X/Fe] but these are not the abundance ratios used to build the tree (see further  discussion in Appendix~\ref{sect:traits}). 

The chemical abundances are examined in four main nucleosynthesis channels \citep{Jofre2020} that represent different epochs after a given star formation episode and different rates of chemical enrichment. The $\alpha$-capture elements of the top panel enrich the ISM through the death of massive stars during the first 100 Myr after a star formation episode \citep{Arnett1978, Tinsley1979}. The iron-peak elements in the second panel from the top originate from explosions of white dwarfs that accrete mass from binary partners \citep{Ptitsyn1980,Kobayashi2011, Nomoto2013}. They start enriching the ISM about a Gyr after the star-formation episode \citep{Tinsley1979,Matteucci2001}. Not all iron-peak elements follow the same trend. 
The odd-Z elements classified in the third panel from the top are formed in shell-burning massive stars, similar to $\alpha$-capture elements, but with a special dependency on metallicity, so their enrichment rate varies with time \citep{Kobayashi2006}.  
Neutron-capture elements in the bottom panel are heavier than iron. They are produced by the rapid($r$)-process (produced from the explosions of massive stars that have been stripped of their outer envelope of hydrogen and mergers of neutron stars) \citep{Argast2004,Wanajo2006} and the slow($s$)-process (produced in asymptotic giant branch stars) and so contribute to the enrichment of galaxies over a range of timescales \citep{Busso1999, Karakas2014,kobayashi2020origin}. 

Stars in magenta and red range in age between 0 and 8 Gyr, are generally on lower-eccentricity orbits, and have low $\alpha-$capture/iron and odd-Z/iron ratios. Therefore, we can attribute them to the thin disk of the Milky Way. Stars in blue are old (8-10 Gyr) and have high $\alpha-$capture/iron and odd-Z/iron ratios. They move on orbits of a range of eccentricities, and therefore we attribute this group to the thick disk, in agreement with previous studies \citep{Nissen2017, Jofre2017, Bedell2018}.  Neutron-capture/iron abundances, in particular those produced through the $s$-process, are more enhanced for the young stars, but those produced mainly through the $r$-process are higher in the thick-disk stars. This is consistent with conclusions from independent chemical evolution studies \citep{Battistini2016}.  We call thin-disk stars with ages comparable to the Sun (magenta) `solar contemporaries' and young stars (red) just `young'. The youngest stars are on the least eccentric orbits as they have had less time for their orbits to have `kinematically' heated through interactions with other stars. 

We call those coloured in yellow `burst' stars. In terms of age, kinematics, and chemical abundances they are intermediate between the thick disk and thin disk, though sometimes follow abundance trends like observed for the thick-disk stars and sometimes like observed for the thin-disk stars. They are therefore unlikely to have originated from outside the Milky Way, which is why they do not appear as a separate branch from the thin and the thick disk branch. We comment on a similar population in our previous work in \citep[see discussion about ``orange population'' in ][]{Jofre2017}. Our method and data to build the tree in that work did not allow us to resolve the branching pattern of the oldest stars (see also  Fig.~\ref{fig:bedelloverlap}a), hence not making it possible to conclude the nature of this population. In our new analysis, which contains more, and more indicative traits, as well as more stars, we see that the orange branch is indeed ancestral to the magenta and red branch. Therefore, we reject the possibility of such stars being of extra-galactic nature \citep[see][for discussions]{Jofre2017}.

The yellow group has a notably higher degree of chemical similarity between them than presents between stars in the other groups, reflected by overall very small error bars in the abundance means plotted in Fig~\ref{fig:abundance_groups}b. This explains the lack of hierarchical structure within them in the tree, as well as the low percentage score of their nodes (See Fig.~\ref{fig:bootstrap} and discussion in App.~\ref{sect:bootstrap}). In their distributions in Fig~\ref{fig:kinematics1} and \ref{fig:kinematics2}, they do not clump in dynamical space.  Therefore, it is unlike they come from a cluster. 

Looking at the distribution of birth radii, they appear to have originated from the inner regions of the Galaxy. However, the thick-disk stars also appear to have originated from further in. This is likely a selection effect. The inside-out formation scenario of the Milky Way disk predicts that iron content of the ISM increases with time and decreases with distance from the Galactic centre \citep{frankel2018}. Therefore selecting stars with an iron content close to that of the Sun biases the sample towards either nearby stars born recently (i.e. the thin-disk stars) or older stars born further in that have travelled to us on eccentric orbits (i.e. the burst and thick-disk stars). This explains why their $z$ component of angular momentum is biased towards smaller radii.

This analysis allows us to conclude that the lower branch corresponds to stars belonging to the thick disk while the upper branch corresponds to stars belonging to the thin disk. 

\begin{figure*}
\hspace{-0.1pt}
\includegraphics[width=0.8\textwidth]{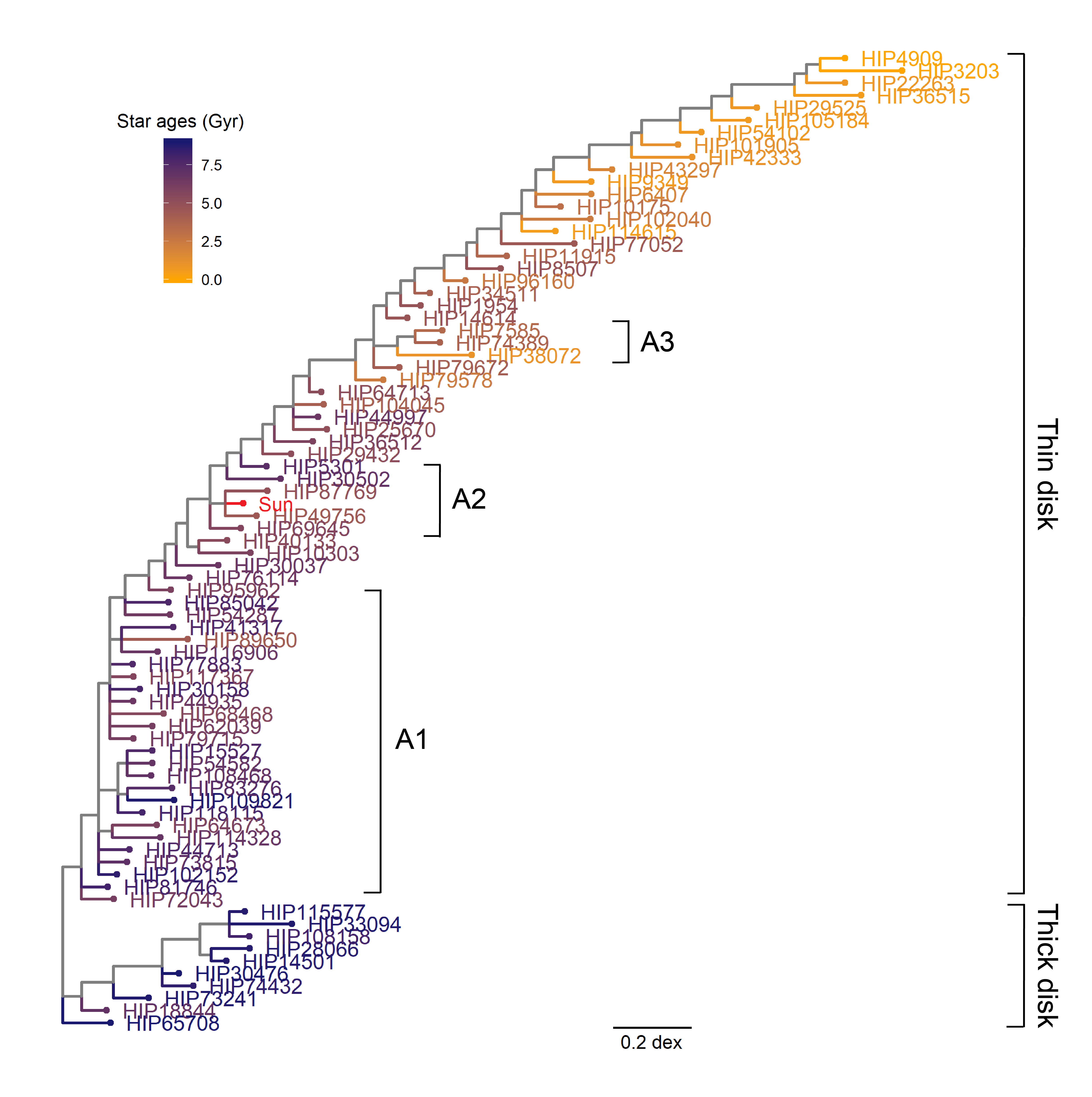}
\caption{Phylogenetic tree constructed from 78 solar twins using 55 evolutionarily significant traits.}
\label{fig:bedell_tree}
\end{figure*}

\subsection{Stellar phylogeny}

Our final tree is illustrated in Fig.~\ref{fig:bedell_tree} , which is the same tree we plotted in Fig.~\ref{fig:abundance_groups}a but here the stars are color-coded by their ages. In the figure, we further include the naming of branches and clades, such that we can proceed with their astrophysical interpretation. In this section, we will focus on what this tree tell us about our chosen sample. 

It is worth clarifying that all livings species share a common ancestor and all are related to one another. Theoretically, it is possible to construct a phylogeny for any combination of plants, animals, fungi or microbes that will reflect the shared evolutionary history of those taxa. However, reconstructing the Tree of Life in its entirety would require sampling not only every living species on the planet but also all the extinct lineages that have ever existed. Thus, biologists have necessarily had to use trees reconstructed from a subset of living and extinct taxa as a proxy for the Tree of Life. The same is true of our stars - the tree we find will only show the branches that link the stars in our sample. Thus the component of galactic history that we reconstruct is determined by the way in which we select our sample of stars. We can choose a sample with a known selection function, but we must ensure that all the stars included are subject to the same quality of analyses so that any differences among them represent real signals of chemical evolution. Alternatively, we can choose a sample whose traits are homogeneously measured and whose uncertainties are equal among measurements, sacrificing the clarity of a known selection function. Regardless of how we chose to select our sample, the choice of stars will impact the final tree, but that does not invalidate the principle of the relationships among the stars in the galaxy. This is further discussed in Appendix~\ref{sect:bed_nis}. 

Our phylogenetic tree shows three interesting features from which we can reveal the evolutionary history of the Milky Way. The first is that our tree is asymmetrical or imbalanced;  the second is that the tree’s general branching pattern appears to correlate with the ages of our sampled stars; and the third is a number of areas on the tree that deviate from that general pattern (A1, A2 and A3 in Fig.~\ref{fig:bedell_tree}). 

\subsubsection{Tree imbalance}
Our tree is asymmetric or imbalanced because the stars are not distributed evenly among its branches. For example, in our tree the 
thin disk contains far more of the sampled stars than the 
thick disk. Imbalanced trees are common in biological systems and are thought to be a consequence of the rate of species formation changing between lineages throughout evolutionary history \citep{Heard1996}. In our case, this is likely a consequence of the unknown selection function. The tree therefore should not be used to e.g. find the relative number of thin and thick disk stars in the Solar Neighborhood by counting stars in each of the branches.  While samples exist from current spectroscopic surveys \citep[see e.g.]{2016MNRAS.460.1131S, 2019A&A...621A..17M, 2020MNRAS.493.2042E} with better understood selection functions, these samples have spectra of significantly lower signal-to-noise and resolution than our sample. That leads to abundance ratios that have higher uncertainties \citep{roederer2014search,adibekyan2016zeta2,Jofre19} and lower number of abundances measured. This could be reflected in a tree that is poorly resolved if the same traits as here were employed. Performing a study using survey data with our current method is beyond the scope of this paper but remains part of future development for our method. 

Even though the solar-twin sample contains stars without a well-understood selection function, our tree’s imbalance suggests that our sample of nearby solar twins are primarily of the thin disk and trace the chemical evolution there until the present day. It is expected that a sample with a better understood selection function and same highly accurate and precise abundance ratios would not be in strong contradiction with our current findings. We know that we live embedded in the thin disk, where chemical evolution is still taking place, and that the majority of nearby stars belong to that population  \citep[e.g.][]{Nissen2020, miglio2020age}. Our phylogeny is consistent with that knowledge. 

\subsubsection{Timeline} \label{sect:timeline}

Our branches, particularly those connecting to stars that are younger than 6 Gyr, have a branching pattern that generally agrees with the ages of the sampled stars i.e. the young stars are close relatives of young stars, the intermediate-age stars are closely related to other intermediate-aged stars, and the old stars are close relatives of other old stars. This age-dependent branching pattern in the thin-disk conforms to our prior expectation that by sampling these stars we are really repeatedly sampling the ISM through time. This is a product of having selected traits that evolve with time. In biological terms, it is like our stars are multiple fossils separated in time but sampled from the same evolving lineage. Thus, some of what appears to be a time-dependent branching process may actually be {\it anagenesis} --  the gradual evolution of a single lineage through time, the ISM (see App.~\ref{anagenesis}). 

In order to further constrain this time-dependent branching process for chemical evolution the stellar sample we have chosen might not be optimal. Firstly, with data we can never be certain how much stars might have migrated from their birth place, confusing us with their actual chemical signatures and positions in the Galaxy \citep{sellwood2002, minchev2019, sharma2020}. Secondly, it is possible that several of these stars are actually siblings (e.g. coeval and with identical chemical composition) but the uncertainties of their ages and chemical abundances measured imply they look different for the tree building method. Thirdly, stars can be doppelg\"angers \citep[having essentially same chemical compositions but being unrelated, e.g., ][]{ness2018}. In fact, we have assumed that every star is sampling a different stellar generation. There are alternative samples of stars that might help us to be more certain about this assumption (for example, open clusters), but that implies developing a different method than here due to the challenges in the homogeneous spectral analyses of open clusters of different ages and metallicities at the precision level of our solar-twin sample for a large sample of clusters \citep{BlancoCuaresma2018, Casamiquela2020}. Alternatively, one could use synthetic data. Comparing models to data is however challenging due to uncertainties in both, models and data \citep{matteucci2012chemical}. Therefore, undertaking such a study is subject of further investigations. 

Nonetheless, the fact that our sampled data shows in the thin disk branch that stars close in age are close in the tree is very encouraging.  Not only because it implies using phylogenetic tools can be used to study relationships of stars and their shared chemical evolution, but because the fundamental idea of chemical tagging for reconstructing the building blocks of the disk \citep{Freeman2002} can indeed work. 

\subsubsection{Anomalies}
There are a number of nodes on the tree where there are deviations from a general time-correlated branching pattern producing clades (see App.~\ref{clade}). Their position on the tree may be linked to a historical event. 
We have  labelled notable anomalies A1, A2 and A3 in Fig.~\ref{fig:bedell_tree}. A1 is the concentration of stars at the bottom of the thin disk branch. The hierarchical differences among these stars are poorly resolved (see also the poor support of these nodes in Fig.~\ref{fig:bootstrap}). This leads to a large polytomy (see App.~\ref{polytomy}). 
 While our stellar sample does not follow a clean selection function, we have no reason to believe that we would be choosing by chance more stars with the properties of A1 among other older or younger solar twins. Thus, A1 may point to an event that increased the rate of star formation at that time, whose implications are discussed further in Sect.~\ref{sect:discussion}.

The Sun is located in A2. Since the ages in that clade are also very similar, the immediate interpretation is that the stars in A2 are the solar siblings from from the same cluster. If that were the case and chemical tagging could work, one would expect their kinematics to support a common origin. However, they do not clump in dynamical space (see below), which suggests they do not come from the same region of the Galaxy. A2 might be yet another signature of enhanced star formation as A1, but less extreme than A1. It could be related to the first significant collision of the Sagittarius dwarf galaxy about 5-6 Gyr ago. Indeed, the recurrent impact of the Sagittarius dwarf on the star formation history of the Milky Way has been characterised with Gaia data \citep{Ruiz-Lara20}, suggesting that the Solar system formation was a consequence of the first of such collisions. In that work, however, the Sun was not part of the sample they studied. Here we include the Sun, and find its place in the tree supports that claim. We must be aware we still have too few stars and interpreting the significance of A2 with a star formation enhancement is highly speculative but illustrates the kind of events we might be able to identify in the Milky Way history with our method.  

A3 refers to the very young star (HIP38072 with an age of 1 Gyr), which is phylogenetically closer to stars that are approximately 3 Gyr older. At present there is no evidence in the literature that this star is binary, and its dynamical behaviour follows the rest of its stellar contemporaries. It might be a blue straggler, or a `straggler-to-be' \citep{mccrea1964extended,hills1976stellar}. These stars gain mass from a companion as a result of binary star interactions. If its mass increases, stellar models of isolated stars predict a younger age than the true age \citep{Jofre2016,yong2016graces}. A follow-up of its radial velocities would be needed to study further its binary nature. 

\begin{figure*}
\centering
\includegraphics[width=\textwidth]{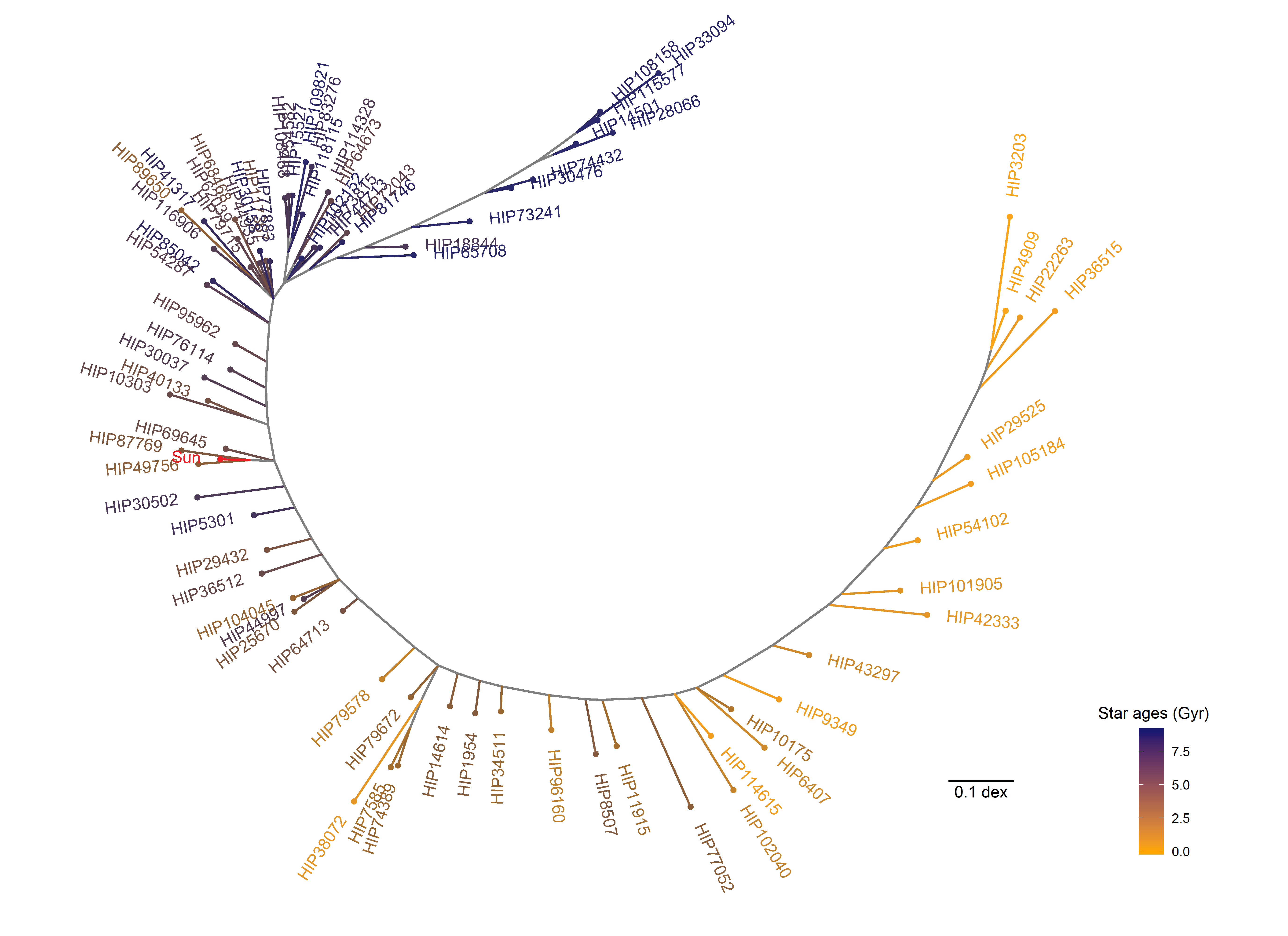}
\caption{Unrooted tree. Same as Fig.~\ref{fig:bedell_tree} but in radiation form.}
\label{fig:tree_radiation}
\end{figure*}

\section{Discussion}\label{sect:discussion}
Here we discuss how we may root our phylogenetic tree and the possible interpretations that emerge from analysing the structure of the tree.

\subsection{The root of the tree}\label{ssect:rooted}
When a tree is rooted we can study the properties of the ancestors in our sample (see App.~\ref{root}). However, not every tree is rooted, making unrooted trees interpretation a bit more challenging. 
While the form chosen to display a tree does not affect its phylogeny, it may affect whether we interpret it to be rooted as this quickly points us towards the beginning of the shared history of our sample. 
The trees in classical form (Figure~\ref{fig:bedell_tree}) might indicate that the thick and thin disks are independent populations that share a common ancestor (instinctively suggested by assuming the root is placed at the left hand side of the tree). However, that tree has no root, and therefore such an interpretation might not be correct. A standard approach to root a tree constructed for biological systems will typically include identifying a species that is distantly related to the species of actual interest -- the so-called ``outgroup". Because it is known \textit{a priori} that all the species of interest are more closely related to one another than any are to the outgroup species, then the point on the phylogeny where the outgroup species connects to the tree will serve as the root. For example, if a phylogeny of birds included a species of crocodile as an outgroup, then the point at which the crocodile tip connected to the rest of the tree would be the root, as all birds are more closely related to each other than any are to crocodiles. In theory, it would be possible to include an outgroup for our stellar phylogeny. We could choose a star that belongs to another component like the halo or the bulge, or a star that is bounded to another dwarf galaxy. However, it would require the analysis of stellar spectra using different techniques to those used to analyse the stars in this sample, introducing new systematic uncertainties \citep{Jofre19} into our distance matrices. 

When our data is plotted in unrooted form (see Fig.\ref{fig:tree_radiation}), an interpretation may rather be that the thick disk is an ancestral population of the thin disk. This interpretation however is based on the stellar ages, which show a directional evolution from thick disk to thin disk.  Our tree in radiation form allows us to reconstruct the shared history of our sampled stars in the context of current scenarios of Milky Way formation and evolution. Here we discuss three such scenarios. 

\subsection{Reconstruction of the galaxy history in the context of mergers}

Several recent studies have discussed the evidence about debris form merging events in the early assembly of the Galaxy \citep[see][for recent review]{helmi2020}. We have learnt that a proto-disk formed rapidly during the first 1-2 Gyr of the Milky Way's existence, namely 8-10 Gyr ago. That rapid formation is coupled with a rapid chemical enrichment rate. 
Hence, the proto-disk contains stars that are enhanced in $\alpha$ to iron-peak abundance ratios, like the old blue stars (see Fig. \ref{fig:abundance_groups}), which we attribute to the thick disk and corresponds to the top right dark branch in the unrooted tree of Fig.~\ref{fig:tree_radiation}. Our tree supports the idea that about 9 Gyr ago, a significant event occurred which altered the evolution of the thick disk and the thin disk. In the absent of such event, we would expect an anagenesis (see App.~\ref{anagenesis}) for the entire sample. Instead, the branching pattern in the tree has a significant polytomy (see App.~\ref{polytomy}) between the thick disk and the thin disk branch, which is labelled as clade A1 in Fig.~\ref{fig:bedell_tree} and is coloured in yellow in Fig~\ref{fig:abundance_groups}a. The thin disk forms and evolves after this polytomy. The event disturbing our tree might be attributed to the Milky Way merging with a smaller galaxy ($\sim10^9 M_\odot$), dubbed the ``Gaia Enceladus-Sausage'' \citep[GES][]{Koppelmann2018,Belokurov2018,Helmi2018}. The system deposited its stars in the Galactic halo that can be observed and studied  \citep[e.g.][]{Das2020}, but are not included in our sample. However, our tree can help us study the effect that such a merger had on the disk. 

Several works have suggested that mergers could have kinematically heated the stars in the proto-disk \citep{Quinn1993,Purcell2010,Minchev2013,Bignone2019, gallart2019birth}. This may have produced the thick disk we observe today \citep{Helmi2018,Belokurov2020}, and paved the way for a new thin disk to form. That would explain how the thick-disk branch appears as an ancestral extant population of the thin disk branch, with a polytomy in between suggesting that an event altered the evolution of the disk. 

The merged system may have also deposited its metal-poorer gas in the Milky Way, which then cooled and eventually settled on the disk and mixed with the pre-existing gas \citep{Brook2007, Bignone2019, Buck2020}. The addition of new gas from GES to the existing gas disk would have altered (and increased) the rates of star formation leading to detectable changes in the chemical traits we used to trace the phylogeny. In fact, the increased rate of star formation could have produced a star formation burst \citep{Bignone2019, grand2020sausage}.
We would however expect a time delay, as the gas needs to cool down before it can form stars. Previous works indicate that such delays are approximately 1-2 Gyr \citep{Rodriguez2019}. The stars in the polytomy have ages that are consistent with forming 1-2 Gyr after the merger of the GES 9 Gyr ago. 

Looking at the stars of the polytomy (in yellow) in Fig.~\ref{fig:kinematics1} and Fig.~\ref{fig:kinematics2}, they do not clump in phase space, like the rest of the stars in our sample. This provides further support to the interpretation that they could be a product of a star formation burst. The homogeneity in the chemical abundances of the burst stars and its spread in dynamical space suggest a very efficient mixing of the GES gas with the pre-existing gas over a large region in the Milky Way, as well a drastic enhancement of star formation \citep{haywood2019}.

The rate of star formation then declines again, but there is sufficient gas to produce stars in the thin disk until the current day.  This is consistent with the age and branching pattern correlation we see for the thin disk, in which the youngest stars are at the end of the branch.

After the Milky Way settled down from the potential merger, the tree supports the idea that the evolution of the Galaxy has been relatively quiescent. Events such as pericentric passages of Sagittarius may have however led to local instabilities \citep{Antoja2018}, driving temporary changes in the star formation rate \citep{Ruiz-Lara20} and therefore the chemical enrichment rate. It is notably interesting that \cite{Ruiz-Lara20} found that the Sun formed in a local star formation enhancement episode because of the interaction with a pericentric passage of Sagittarius. Our tree supports that finding because the Sun is located in a smaller polytomy (A2). However, we need more stars and a better understanding of the sample selection function to properly link the A2 clade to the conclusions of \cite{Ruiz-Lara20}. 

\subsection{Reconstruction of the galaxy history in the context of gas infall}

A classical explanation to the observed discontinuity of [$\alpha$/Fe] abundances for the thin and the thick disk stars is the two-infall model of chemical evolution \citep{Chiappini1997}. There, the disk formed in two episodes which depend on the timescales of gas infall. The first episode is characterised by a short-period gas infall whose duration is of the order of 1 Gyr depending on the study \citep[see recent discussion in e.g.][]{spitoni2020}. The second episode is characterised by a significantly longer period (longer than the age of the Milky Way) of slower accretion of gas \citep{miglio2020age}. This leads to a rapid formation of the thick disk, and a slower formation of the thin disk. The two-infall model also predicts a quenching of star formation between both accretion episodes. That quenching causes a discontinuity in star formation, and a significant mix of the in-situ disk gas with the metal-poor infalling gas, which ultimate leads the thin disk to start forming stars with lower metallicities than the youngest stars from the thick disk. 

Our tree shows two independent branches for the thin and the thick disk, in which the thick disk could be an ancestral population to the thin disk. There is however an abrupt divergence between the thick and thin disk, suggesting that there was a chemical discontinuity in the transition of the thick to the thin disk. The two-infall model predicts this discontinuity.  Moreover, looking at the ages of our stars, we indeed see that there is a difference between the youngest  and oldest stars of the thin and thick disk, respectively (see online table). This is consistent with the predictions of the delay between the infall episodes of some of the the two-infall  models. The age difference in the stars of our branches, however, is smaller than the predictions \citep{bensby2003elemental,feltzing2003signatures,kilic2017ages}. We recall that our stellar sample does not have a known selection function to make it suitable for direct comparisons with models in this way. 

Between the thin and thick disk branches there is the A1 polytomy which contains stars of ages in between the thin and the thick disk.  We might relate this polytomy to the `loop' in the [$\alpha$/Fe]-[Fe/H] diagram \citep[see e.g. Fig. 6 of][]{spitoni2020}. This loop is produced due to a `bump' and a `drop' in the metallicity and [$\alpha$/Fe] of stars that formed 6-8 Gyr ago, which correspond to the ages of the stars from our polytomy. These features are signatures of the delayed gas infall in the two-infall model. \cite{spitoni2020} commented that  the presence of such features is not obvious in the observations.  Perhaps using an evolutionary tree to compare the two-infall model and observations offers a way to better identify the presence of such features.  

\subsection{Reconstruction of the galaxy history in context of radial migration}


Radial migration refers to stars that have moved from their birth place through interactions with bars, spiral arms, or orbiting stellar systems \citep{sellwood2002}. While there is a general consensus that radial migration is taking place in the Milky Way, its impact on key observables, such as the chemical distinction of the thin and the thick disk in their [$\alpha$/Fe]-[Fe/H] distribution, remains under debate \citep[e.g.][]{minchev2019, haywood2019, sharma2020, Buck2020}. Quantifying the importance of radial migration has been difficult mostly because of the lack of good kinematic data to constrain dynamical models of the disk, but this is now rapidly changing thanks to the wealth of accurate data of kinematics from Gaia complemented to chemical data from spectroscopic surveys. New results present evidence that stars indeed have moved few kpc from their birth places \citep{frankel2018, minchev2019}, which we see in our data as well (see Fig.~\ref{fig:kinematics2}).

The basic principles of chemical evolution and cosmological models of Milky Way formation we have used in the two scenarios above to interpret our tree have however neglected radial migration. That is, we have interpreted the branching pattern as the cloud evolution from which stars have formed, died, and passed-on their chemical make-up, and have not accounted that stars might be tracing the chemical evolution of other clouds because they have moved from their birth places. Without considering radial migration, we might be affected by a kind of Yules-Simpson paradox in Galactic archaeology \citep{minchev2019}. Without considering the information of the stellar birth radius it is difficult to draw conclusions about the evolutionary history of the Solar Neighborhood since a trend seen in stellar age might be significantly affected when considering the birth radii. 

We clearly have stars from different birth radii (see Fig.~\ref{fig:kinematics2}) and therefore it is not obvious what is the relation of the branching pattern with time in our tree. Moreover, the birth radii become increasingly uncertain with stellar age, 
for it is known that stars do not retain much dynamical memory after 1 Gyr. Therefore, our interpretations come from basic principles and consider the star formation happening homogeneously through the entire thin and thick disk. This is of course an over simplification of reality - there is enough evidence pointing towards an inside-out formation of the disk \citep{frankel2018, frankel2020, Buck2020} producing a dependency of star formation efficiency (hence chemical enrichment rate) as a function of Galactic radius. This is reflected  in significant metallicity gradients across the disk \citep{magrini2017}. 

Our branching pattern might be a result of a combination of radial migration and chemical evolution. To disentangle between both we need to apply our phylogeny tools on models considering both, dynamical and chemical evolution. So far, chemodynamical modelling studies are focused on $[\alpha$/Fe] and [Fe/H], and most of the publicly available chemical evolution models with several elements do not include radial migration \citep{spitoni2015effect}. Therefore, undertaking such a study remains part of future work.  

However, our tree shows two aspects that are worth highlighting. The first aspect is the notable break between the thin and thick disk branches, which challenges a smooth transition between both populations in the context of radial migration, although this could be mitigated considering the time delay of SNIa \citep[e.g.][]{sharma2020}. The second aspect is the polytomy A1. It would be interesting to understand why stars of 6-8 Gyr, which are very chemically homogeneous but have a range of orbits and birth radii, have a preference to migrating more to the Solar Neighborhood than stars of other ages with similar iron contents but different chemical abundances.

We finally remark that a possible Yules-Simpson paradox here \citep[namely that we might be confusing the evolutionary relationships of our stars with radial migration][]{minchev2019} is a common challenge in other phylogenetic studies. Migration is not only happening in the Galaxy, it happens everywhere. Building models and tools to deal with disentangling both effects (direct hereditability in a population and exchange of traits between populations) is on-going research \citep[see e.g.][]{LIU2009320}. Therefore, the Yules-Simpson paradox in our case should not be seen as a limitation of our method but an opportunity to provide new ways of disentangling the effects of both time and dynamics in the evolutionary processes affecting the Milky Way. 



\section{Concluding remarks}\label{sect:conclusions}
Here we present a robust methodology using methods in evolutionary biology that effectively visualizes high-dimensional chemical abundance data in terms of evolutionary relationships between stars. Using phylogenetic methods allow us to reconstruct a history for our Galaxy which agrees remarkably well with findings in recent independent works. It is expected that using more stars for which we have a better understanding of the selection function will help us to better constrain current models of galaxy formation and evolution.

All evolutionary problems, be they biological or astrophysical, involve pattern and process. Phylogenetic trees are an excellent means of discovering patterns, from which the processes driving the evolution can be inferred. It is currently uncertain how chemical information is passed through time, or, in effect, from one generation of star formation to another. However, assuming there is a level of heritability in chemical structure, we have been able to use this to apply phylogenetic techniques to discover a pattern of evolution in the Galaxy. We show how the diversity of chemical compositions in nearby stars can be used to shed light on their shared history. The phylogeny we find from examining just chemical abundances of a relatively small sample of stars agrees remarkably well with the evolutionary history inferred from the analysis of both simulated data and much larger samples of photometric, spectroscopic, and astrometric data.

Phylogenetics is a powerful way of organizing and visualising the data and applying it in this new context can help us to understand the evolution of the galaxy in which we live. Through it we can identify and study significant events that shaped the Milky Way until it formed our Solar system, and work our way to learning more about the process through this depiction of the pattern. Although the mechanisms of biological and stellar evolution are entirely different, the fact that the two have a mechanism of heritability allows the same approaches to be applied. Since chemical elements are inherited between stellar generations, we have been able to find the chemical signatures of evolution in the Galaxy.

Darwin’s thinking, originally applied to biological systems and the origin of species, has been extended to the history of cultures, languages, technological systems and even religions \citep{Gray2009, Retzlaff2018, Jetz2012, Upham2019}. In each, the process is different, but the existence of a form and level of heritability allows us to discover history and infer the process. We have extended it here to Galaxy evolution. It is likely that developing better models of how stellar abundances change their heritable signatures, and incorporating more stars, will reveal not just a better understanding of the Galaxy, but also how our Sun inherited the chemical composition suitable for a planet on which life can evolve.

\section*{Data Availability}
The data used for this paper is submitted as an online table which will be published via CDS with the paper if accepted.

\section*{Acknowledgements}
H.J. acknowledges support from the MIT International Science and Technology Initiatives (MISTI) grant, which funded her research visit to UDP. P.J. acknowledges financial support of FONDECYT Iniciaci\'on grant Number 11170174 and FONDECYT Regular grant Number 1200703.  P.D. acknowledges funding from the UK Research and Innovation council (grant number MR/S032223/1).


\bibliographystyle{mnras}
\bibliography{references2}

\appendix
\section{Biology glossary and concepts}
In this appendix we describe a few key concepts that serve to transfer the field of phylogenetics into galaxy evolution. These concepts are used throughout the paper. 

\subsection{Descent with modification}\label{descent}
 Some characteristics are passed from one generation to the next. Since the passing is not perfect, some characteristics are modified. With time, these differences accumulate. Descent with modification creates hierarchies of similarities, such that close relatives will be more similar in many respects. To uncover evolutionary relationships, a key challenge is to distinguish characteristics that are similar by descent from those that have evolved independently. 

\subsection{Speciation}\label{speciation}
The formation of new and distinct species in the course of evolution. No species definition provides an unambiguous way to delineate all groups, as traits might vary in their level of dissimilarity. Since differences accumulate continuously, there is no clear line that separates a species from its relatives or its ancestors.  

\subsection{Taxa}\label{taxa}
Taxa refer to the groups we wish to study in a phylogenetic tree. They can be groups within a species, or different species. We take their traits and compare them to find their similarities and relationships, in order to use the branching pattern to reconstruct their shared history. 

\subsection{Root}\label{root}
Since phylogenetic trees visualise the similarities and differences between a set of chosen traits, they do not necessarily have a root. This is particularly the case in trees constructed with the distance method, which was employed in this paper.  Unrooted trees are solely branching diagrams that show the relationships between traits. On the other hand, rooted trees identify the first common ancestor, and therefore indicate the direction of descent.  

\subsection{Clade}\label{clade}
Part of a phylogenetic tree that includes an ancestral lineage and all the descendants of that ancestor. 

\subsection{Polytomies}\label{polytomy}
Polytomies occur when the signal of the branching pattern is lost and branches split into more than two branches. There are two kinds of polytomies: soft, which are a consequence of the data having insufficient resolution to discern the correct branching patterns, and hard, which are actual instances where a single evolving lineage has diversified into three or more separate lineages \citep{Lemey2009}. In other words, hard polytomies would show in a tree when three or more lineages have at some point in history exact trait difference among them. In biological systems hard polytomies are thought to be exceptionally rare \citep{Lemey2009}. However, in stellar systems, it is unclear whether hard polytomies are as exceptional, especially in the context of major mergers and star formation bursts in galaxy evolution and  the uncertainties of astronomical data measurements.

\subsection{Anagenesis}\label{anagenesis}
The gradual evolution of a single lineage through time. Traits differences can therefore be traced as information passed through generations but they still represent the evolution of the same population. 

\section{Tree's systematics}
\begin{figure*}
    \centering
    \includegraphics[width=\textwidth]{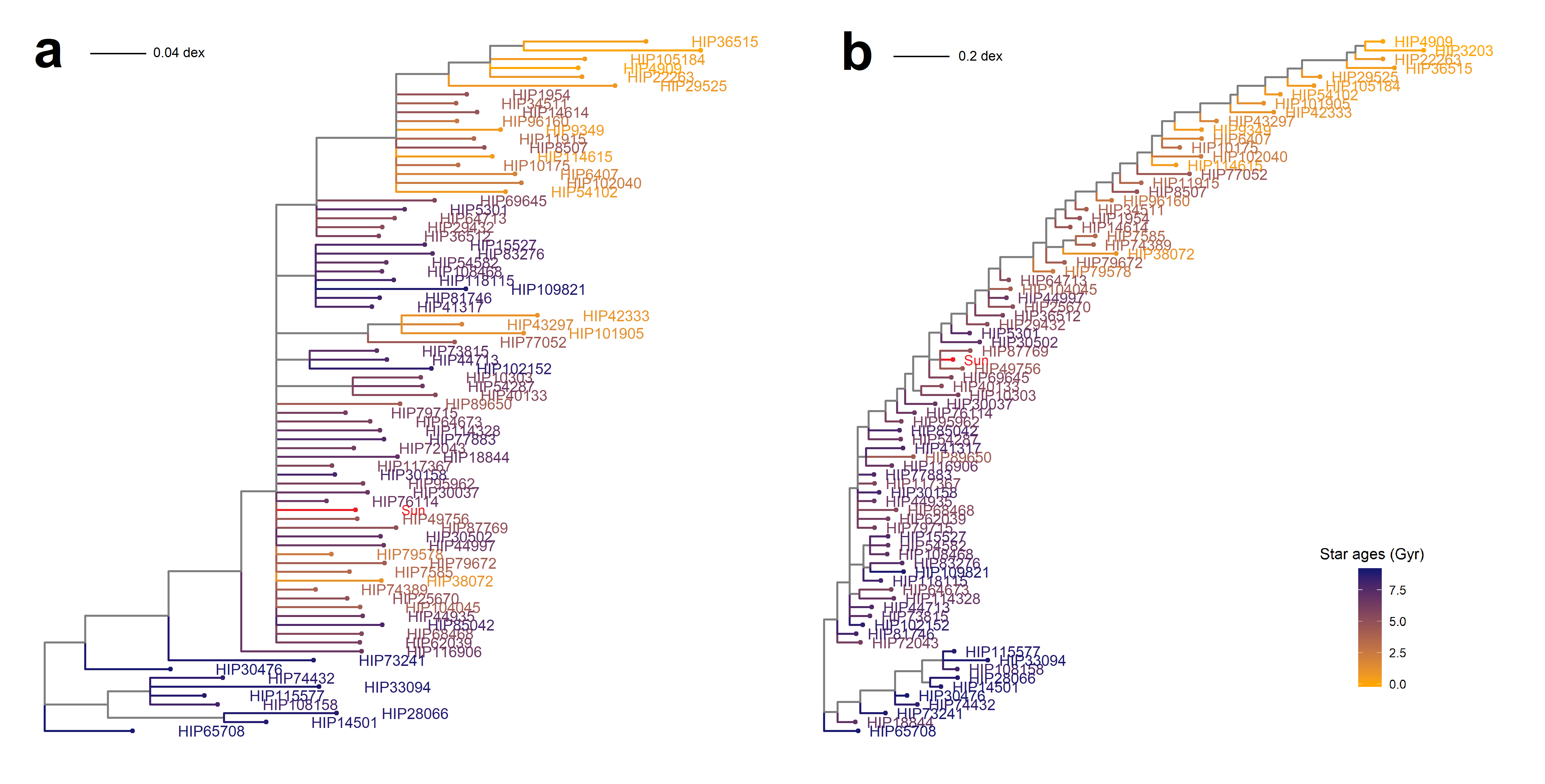}
    \caption{(a) Phylogenetic tree constructed from 79 solar twins using the 30 abundance ratios [X/Fe] as published in \citet{Bedell2018}, (b) Phylogenetic tree constructed from the same stars  but using 55 abundance ratios [X$_i$/X$_j$] from online table.}
    \label{fig:comparetrees}
\end{figure*}

\begin{figure*}
    \centering
    \includegraphics[width=\textwidth]{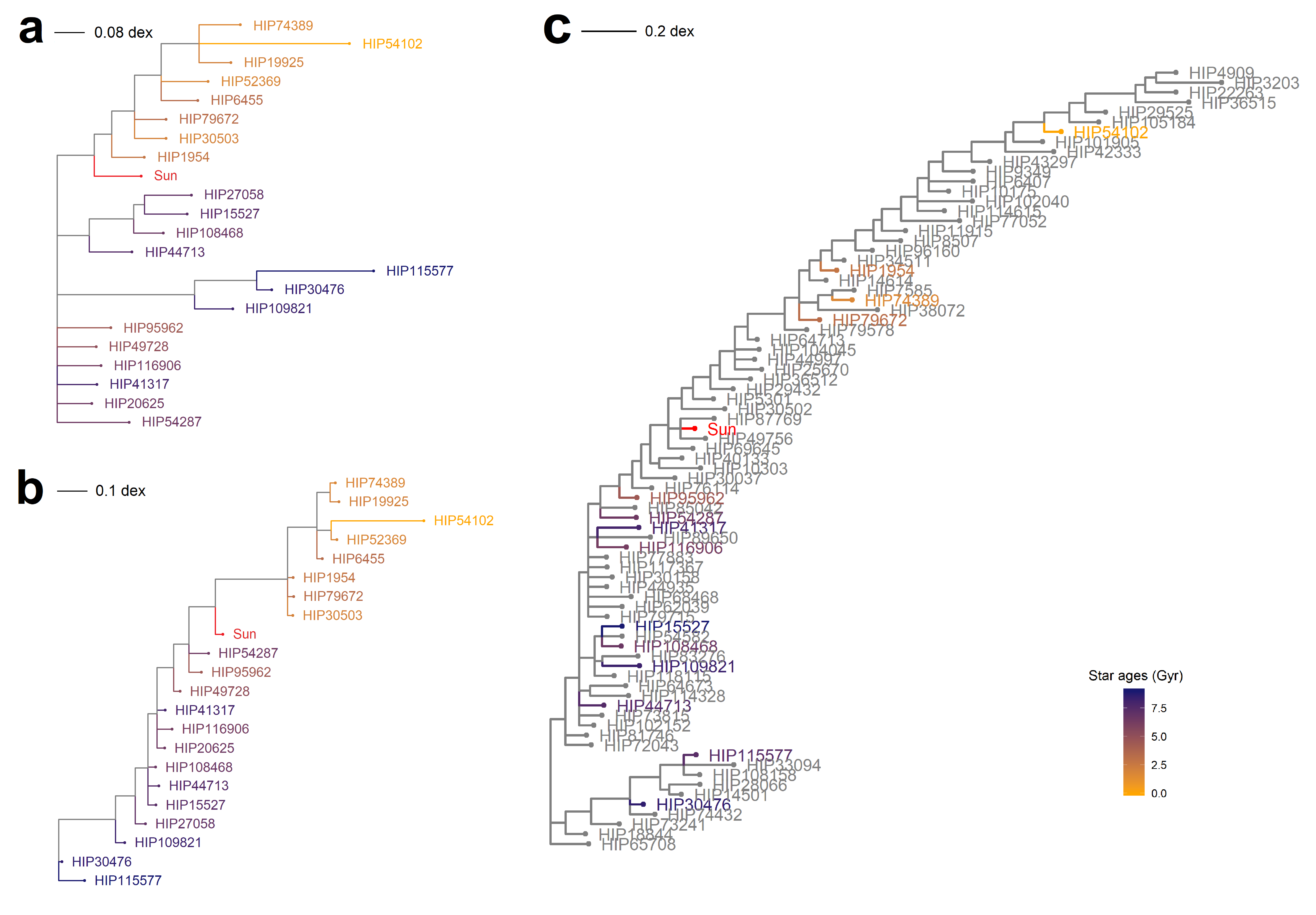}
    \caption{(a) Phylogenetic tree constructed from 22 stars using traits over iron  \citep[from][]{Jofre2017}, (b) Phylogenetic tree constructed from 22 stars using 17 evolutionarily significant traits, (c) Phylogenetic tree constructed from 79 stars using 55 evolutionarily significant traits, 15 overlapping stars from (a) and (b) highlighted}
    \label{fig:bedelloverlap}
\end{figure*}

\subsection{Choosing traits}\label{sect:traits}

The effect of the selection of evolutionarily informative traits over standard abundance ratios [X/Fe] can be seen in Figure \ref{fig:comparetrees}. There we constructed trees using our dataset following the procedure explained in Sect.~\ref{sect:methods} using two different set of traits. The first one results in a tree shown in Fig.\ \ref{fig:comparetrees}a, in which our traits are the 30 elemental abundances over iron [X/Fe] directly measured in \cite{Bedell2018}. Using abundances as a function of iron is typical of much prior work on chemical evolution, and is what we considered in \cite{Jofre2017}. Inspired by the recent works on chemical clocks  \citep{Nissen2015}, and in particular the findings of \cite{Jofre2020} about the abundance ratios that can be used to study chemical evolution, in Fig.\ \ref{fig:comparetrees}b, we show the tree constructed using the 55 traits indicated in Sect.~\ref{sect:data}  (see also Table of online material). Both trees have the stars color coded by their ages, also indicated in that table.  Branch lengths represent the total chemical differences, whose scale can be seen in the scale bar. Note that both trees are differently scaled.

The difference in the branching pattern between Fig.\ \ref{fig:comparetrees}b and Fig.\ \ref{fig:comparetrees}a is stark.  While the base structure of the trees is similar, the tree in Fig.\ \ref{fig:comparetrees}b is significantly more resolved than that of Fig.\ \ref{fig:comparetrees}a.   However, both trees have two major clades (see App.~\ref{clade}), whose astrophysical interpretation is discussed in Sect.~\ref{sect:stellar_pops}. The lower clade contains in general older stars at the base of the tree. The upper clade contains the majority of the stars in both trees, and has stars of all ages, with the youngest stars in general at the top of the tree.

In the tree in Fig.\ \ref{fig:comparetrees}b, we are able to reconstruct with higher resolution the branching pattern and therefore the evolutionary history of the stars on the majority of the tree. In fact, the upper clade in Fig~\ref{fig:comparetrees}a presents a very large polytomy close to its base and several smaller polytomies at different levels towards the top. These polytomties reflect the level of similarity of in many of the [X/Fe] abundance ratios in the sample \citep[see e.g. Figure 3 of][]{Bedell2018}. This suggests that the polytomies in this tree are soft, i.e. the data does not contain enough information to obtain an evolutionary signal.  The Sun, highlighted with red colour,  is located at the large polytomy.   While the tree with the optimized traits (Fig.~\ref{fig:comparetrees}b) still lacks resolution at the base of the upper branch, this opens doors for interesting analysis which we discuss in Sect.~\ref{sect:discussion}.  In that tree, polytomies are concentrated in a small section of the tree but a large number of stars which were in the previous polytomy are now resolved, including the Sun. We are now more confident in idenfitying which stars are more similar to the Sun, which in addition are very close in age.  This  result shows us that  phylogenetic signal might be lost in the noise of traits that are not evolutionarily informative or do not differ significantly among taxa. 

Choosing the right traits is therefore a fundamental part of the analysis. Finding them should further motivate to improve the theoretical understanding of how and why these abundance ratios information is inherited between stellar populations.

\begin{figure*}
\centering
\includegraphics[width=\textwidth]{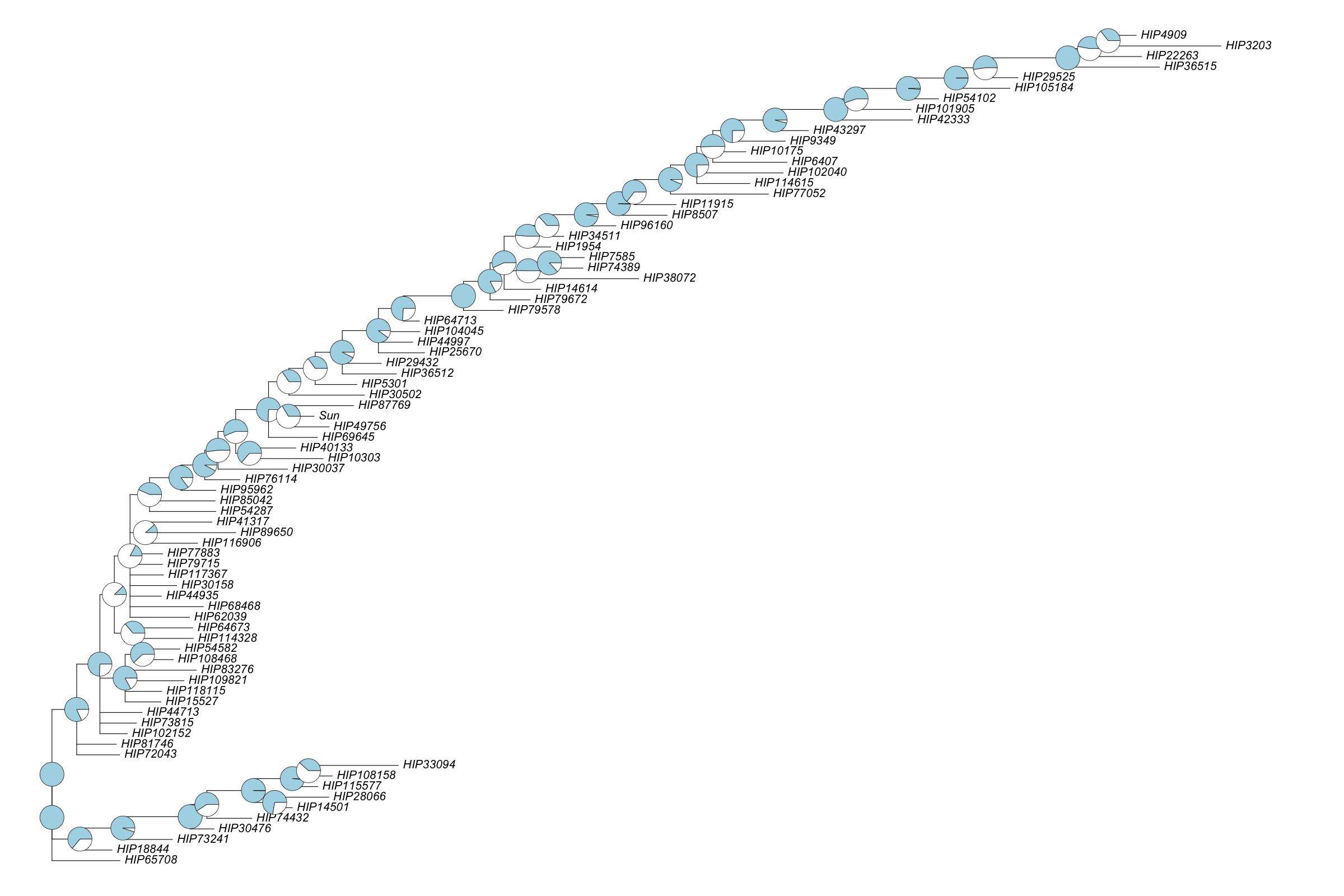}
\caption{Bootstrap support for stellar phylogeny. Pie charts show the percentage of support (blue) for the associated
node among the bootstrapped trees.}
\label{fig:bootstrap}
\end{figure*}

\begin{figure}
\hspace{-0.3in}
\includegraphics[width=3.8in]{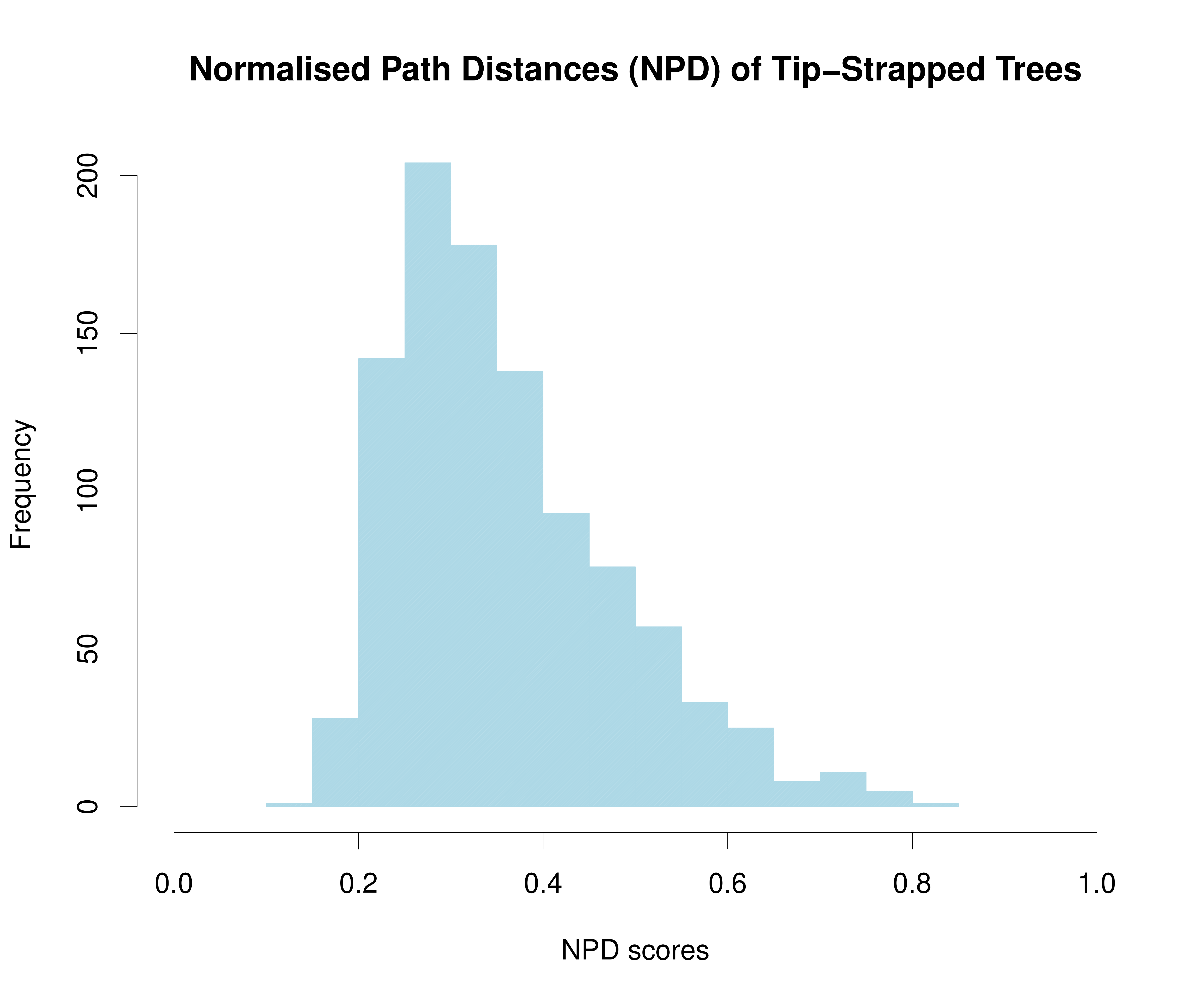}
\caption{ Frequency distribution of the Normalised Path Distance (NPD) scores for the thousand iterations of the tip-strapping procedure. All scores are bellow one, meaning that in all runs the pruned tree and the subsampled trees were more similar to one another than a pair of randomly generated trees.}
\label{fig:tip-strap}
\end{figure}

\subsection{Comparing datasets}\label{sect:bed_nis}
In addition to our dataset, we compared the results with respect to our previous work, in which we used a sample studied by \cite{Nissen2015, Nissen2016} and contains 22 stars, including the Sun, for which abundances for 17 different chemical elements were measured, in addition to ages. Both analysis use very similar techniques and same spectra. Our main sample is bigger in two ways: it contains more stars and more elements. There are 15 stars in common which allow us to make comparisons. 

We used the approach presented in \cite{Jofre2020} to select 17 traits to produce a tree for the stars in the Nissen dataset: [C/Y], [C/Ba], [O/Y], [O/Ba], [Mg/Y], [Mg/Ba], [Al/Y], [Al/Ba], [Si/Y], [S/Y], [S/Ba], [Ti/Y], [Ti/Ba], [Zn/Y], [Zn/Ba], [Sc/Ba], [Cu/Ba]. Because there are stars in common between both datasets, this analysis helps to test the consistency between results as well as between this work and \cite{Jofre2017}.  

Figure \ref{fig:bedelloverlap} compares trees created using different datasets. For reference, in Panel a we show the phylogenetic tree created from the 22 stars from the Nissen sample published in \cite{Jofre2017}. The traits employed in that work were all 17 abundance ratios in the form of [X/Fe] ]\citep[see][for details]{Jofre2017}.  In Panel b we show the same dataset but using the new sets of traits indicated in Sect.~\ref{sect:data}, which are abundance ratios as a function of $s-$process abundances. Additional differences between both trees is the procedure to build them, which has changed from \cite{Jofre2017} to the procedure described in Sect.~\ref{sect:methods}. Panel c shows shows the phylogenetic tree created from the 78 stars in the  sample with the 55 hereditable traits. This tree is the same as in Fig.~\ref{fig:comparetrees}b and Fig.~\ref{fig:bedell_tree}. The 15 overlapping stars are highlighted based on their colors in Fig.~\ref{fig:bedelloverlap}a-b for reference.

Similar to our result comparing traits above, we see how the tree in Fig.\ \ref{fig:bedelloverlap}b is more resolved than the \cite{Jofre2017} version displayed in Panel a, although both trees show a similar behavior of stars of similar ages close in the tree. This indicates that the phylogenetic signal is in the data, but a choice of better traits helps increasing that signal. In particular, in \cite{Jofre2017} we were unable to conclude on the ancestral populations of the Sun's lineage because the tree was not resolved enough for older stars.  With our new procedure the situation improves. More chemical abundance ratios and more stars however improves the situation even further, as we see in Fig~\ref{fig:bedelloverlap}c.  

This comparison allows us to analyze the changes in phylogenetic topology that occur when we increase the number and type of traits used.  However, the addition of traits that include neutron-capture elements among others used to generate the tree in Fig.\ \ref{fig:bedelloverlap}c have increased the resolution. Stars that were in polytomies in Fig.\ \ref{fig:bedelloverlap}b (see HIP108468, HIP41317, HIP116906, HIP79672, and HIP1954) have been resolved in subclades in Fig.\ \ref{fig:bedelloverlap}c.  While the placement of some of the stars has changed between Fig.\ \ref{fig:bedelloverlap}a and Fig.\ \ref{fig:bedelloverlap}c, their grouping as in the main clades has remained consistent between the two trees, as well as the overall dependency of position in tree with the age. 

The test shows that, while trees differ when different datasets are used, the phylogenetic signal remains if it exists. This supports our postulation that stars in the solar neighborhood are related through their chemical composition (even if their common ancestry was far in the past). 

\subsection{Phylogenetic signal in data}\label{sect:bootstrap}


To ascertain confidence values for each node in our phylogeny we applied a further bootstrapping procedure \citep{Felsenstein1985}. This technique is commonplace when evaluating support for phylogenies derived from biological datasets and was also employed before in stellar datasets \citep{Jofre2017}. Specifically, we randomly sampled our alignment of chemical elements with replacement, until we had a new alignment, equal in length to the original (55 traits). We then parse that new resampled alignment to the tree building process described above. We do this a thousand times, and create a distribution of trees derived from these bootstrapped samples. For every node in our original tree, we count the number of times it is represented in this distribution and take this value as a measure of support. We stress that here we did not consider these sample trees for the maximum clade credibiliy tree (see above) but for studying the support in each node of our final tree. 

 The support is illustrated in the nodes of the tree in Fig.~\ref{fig:bootstrap} as percentage pie charts. While the support for specific nodes ranged considerably, from 100\% for six nodes to as low as 11.6\% for one, 40 of the 54 internal nodes showed support above 50\% and 29 above 70\%. We note that a number of our most poorly supported nodes are centred at the root of the top branch, indicating the poor hierarchical pattern of differences. The implications of this are discussed in more detail in Sect.~\ref{sect:discussion}. It is important however to comment here that a poor support for a clear branching pattern is not necessarily a problem. It can well suggest that perhaps the poor resolution relates to these samples being taken from a period of rapid chemical evolution and star formation 
 
 There is no consensus on how bootstrap values for nodes should be interpreted \citep{Soltis&Soltis2003}, however \cite{Hillis&Bull1993} suggest that bootstrap values are a conservative estimate of the probability that the node is real, and that when trees are symmetrical and evolving under equal rates bootstrap values of 70\% or greater correspond to a 95\% or greater probability that the clade is real. While it remains to be seen if those assumptions from biology hold for our stellar data set, the fact that a majority of nodes show greater than 70\% support indicates that the topology is relatively robust to permutation. From this we conclude that we are detecting signal of phylogeny using these chemical abundance ratios.

We also wanted to assess the robustness of our procedure with regards to our choice of stars. To do this we reran the tree building procedure but with a subsample of stars of size $n$, a process we hereafter refer to as tip-strapping. To allow for comparison, we altered our original tree, removing any tips that were not included in the subsample. This left us with two trees with $n$ tips, the pruned version of the original tree which retained the overall structure of our stellar phylogeny, and a subsampled tree. 

To quantify the difference in topology between these two trees, we calculated their Normalised Path Distance (NPD) as described in  \cite{NaserKhdour2019}. NPD calculates the quadratic-distance of our two topologies and normalizes it by comparing it to the average quadratic-distance of a pair of randomly generated trees of $n$ size. This has two advantages. Firstly, it means that even though the size of $n$ varies between iterations of the test, the NPD scores are still comparable as they have been normalised against an expected null-value. Secondly, this normalized value is easy to interpret.  An NPD score of zero means the subsampled and pruned trees are identical, while an NPD scores of one means that the two trees are as dissimilar as the average of two randomly generated trees. An NPD score greater than one means that the pruned and subsampled trees are more dissimilar than the average of two randomly generated trees. We repeat this process 1000 times to produce a distribution of NPD scores. It is important to remark that the left skew of the NPD values is a consequence of the fact that our results are  consistent and that it is almost impossible to score a zero with this test.

Figure~\ref{fig:tip-strap} shows the distribution of NPD scores for our realisations. The NPD scores ranged from 0.149 to 0.819, with a median value of 0.333. Thus, in all cases the subsampled and pruned trees were more similar to one another than two randomly generated trees of equal size. This suggests that despite our subsampling procedure, our tree building process was returning trees that generally reflected the topology of the phylogeny built using the complete sample of stars. This is consistent with our results from the comparison between the trees obtained with the datasets presented in Fig.~\ref{fig:bedelloverlap}.

\end{document}